\documentclass[aps,prl,reprint,unsortedaddress,showpacs,floatfix,nofootinbib]{revtex4-1}
\usepackage[colorlinks=true,breaklinks=true,hypertex]{hyperref}

\usepackage{graphicx}
\usepackage[normalem]{ulem}
\usepackage{amsmath,amstext,amssymb}
\usepackage{dcolumn}
\usepackage{bm}

\newcommand{\mic}{~{\rm \mu m}}

\newcommand{\kpp}{k^{(2)}}
\newcommand{\Iin}{I_{\rm in}}
\newcommand{\imm}{~{\rm mm^{-1}}}

\newcommand{\tw}{~{\rm TW/cm^2}}

\begin{document}


\title{Ultrafast and octave-spanning optical nonlinearities from strongly phase-mismatched cascaded interactions}

\author{B.B. Zhou$^1$, A. Chong$^2$, F. W. Wise$^2$ and M. Bache$^1$}\email{moba@fotonik.dtu.dk}
\affiliation{$^1$DTU Fotonik, Dept. of Photonics Engineering,
  Technical University of Denmark, 2800 Kgs. Lyngby, Denmark\\
  $^2$Dept. of Applied and Engineering Physics, Cornell University, Ithaca, New York 14853, USA}

\date{\today}

\begin{abstract}

Cascaded nonlinearities have attracted much interest, but ultrafast applications have been seriously hampered by the simultaneous requirements of being near phase matching and having ultrafast femtosecond response times.
Here we show that in strongly phase-mismatched nonlinear frequency conversion crystals the pump pulse can experience a large and extremely broadband self-defocusing cascaded Kerr-like nonlinearity. The large cascaded nonlinearity is ensured through interaction with the largest quadratic tensor element in the crystal, and the strong phase-mismatch ensures an ultrafast nonlinear response with an octave-spanning bandwidth. We verify this experimentally by showing few-cycle soliton compression with noncritical cascaded second-harmonic generation: Energetic 47 fs infrared pulses are compressed in a just 1-mm long bulk lithium niobate crystal to 17 fs (under 4 optical cycles) with 80\% efficiency, and upon further propagation an octave-spanning supercontinuum is observed.  
Such ultrafast cascading is expected to occur for a broad range of pump wavelengths spanning the near- and mid-IR using standard nonlinear crystals. 
\end{abstract}

\pacs{42.65.Re, 42.65.Tg, 42.65.Ky, 42.65.Sf}
\maketitle


Ultrashort laser pulses with duration of a few optical cycles are fascinating 
because they are brief enough to resolve temporal dynamics on an atomic level like chemical reactions, molecular vibrations and electron motion.
Additionally since they are very broadband and can become extremely intense, few-cycle pulses are used for coherently exciting and controlling matter on a microscopic level \cite{brabec:2000,*krausz:2009}. However, bandwidth limitations in the amplification process means that only multicycle energetic pulses are delivered from laser amplifiers, which necessitates efficient external compression schemes \cite{DeSilvestri:2004}: the bandwidth needed to support a few-cycle pulse is generated by nonlinear processes, typically self-phase modulation (SPM) from a cubic Kerr nonlinearity. The nonlinear step leaves a strongly chirped and stretched pulse, and temporal compression is achieved by subsequent dispersion compensation (e.g., by gratings or chirped mirrors).

Access to ultrafast and extremely broadband self-defocusing nonlinearities is very attractive for femtosecond nonlinear optics. First, whole-beam and small-scale self-focusing problems encountered in Kerr-based compressors are avoided so the input pulse energy is practically unlimited \cite{liu:1999}. 
Secondly, with solitons both spectral broadening and temporal compression can occur in a single nonlinear material; solitons are stable nonlinear waves that exist as a balance between nonlinearity and dispersion. For a self-defocusing nonlinearity soliton formation requires normal (positive) dispersion, which is predominant in the near-IR where the majority of lasers operate.

Self-defocusing nonlinearities can be achieved with cascaded harmonic generation. This occurs when a frequency conversion process, like second-harmonic generation (SHG), is incomplete due to absence of phase matching $\Delta k\neq 0$, and instead the pump experiences a Kerr-like nonlinearity $n_{\rm casc}^I\propto -d_{\rm eff}^2/\Delta k$, controllable in both magnitude and sign through $\Delta k$ \cite{Ostrovskii:1967,*Thomas:1972,*desalvo:1992}.
Cascaded SHG has been used for high-energy femtosecond pulse compression 
\cite{liu:1999,ashihara:2002,moses:2006,ashihara:2004,*Zeng:2008,*moses:2007,bache:2007a,*bache:2008,bache:2010}, but has not found significant use to date. 
One reason is that ultrafast cascading is limited by the
simultaneous requirements of being near phase matching and having ultrafast femtosecond response times, 
which limits the usable input wavelengths and nonlinear crystals \cite{liu:1999,ashihara:2002,moses:2006,ashihara:2004,*Zeng:2008,*moses:2007,bache:2007a,*bache:2008,bache:2010}. 
Therefore only multicycle compressed pulses were observed in $\beta$-barium borate (BBO) crystals at 800 nm \cite{liu:1999,ashihara:2002} and at longer wavelengths (1260 nm) few-cycle compression was observed only in a narrow operation regime \cite{moses:2006}. 


Here we show a surprising solution that has always been there but has been overlooked for good reason: when exploiting interaction with the largest $\chi^{(2)}$ tensor components, strong and octave-spanning self-defocusing cascaded nonlinearities can be obtained even with huge phase mismatch and group-velocity mismatch (GVM). %
We exploit values of $\Delta k\simeq 500\imm$ and $d_{12}=-500$ fs/mm that are an order of magnitude larger than previous experiments. Ordinarily, this type of interaction is accessed only by quasiphase matching (QPM), providing an indication of the surprisingly large $\Delta k$. 
We demonstrate experimentally the utility of this process by compression of energetic 47-fs infrared pulses to 17 fs (4 cycles of the electric field) through excitation of self-defocusing temporal solitons. This occurs in an only 1 mm long ordinary lithium niobate crystal. Upon further propagation an octave-spanning supercontinuum is observed. 

In cascaded SHG the frequency conversion from the fundamental wave (FW, $\omega_1$) to its second harmonic (SH, $\omega_2=2\omega_1$) is not phase matched: after a coherence length $\pi/|\Delta k|$ only weak up-conversion to the SH occurs, followed by the reverse process of back-conversion after another coherence length. On further propagation this cascade process is cyclically repeated and the FW effectively experiences a Kerr-like nonlinear refractive index change $\Delta n=n_{\rm casc}^I I$ proportional to its intensity $I$. In the strong phase-mismatch limit 
$n_{\rm casc}^I\simeq  -2\omega_1 d_{\rm eff}^2/c^2\varepsilon_0n_1^2n_2 \Delta k$ \cite{desalvo:1992}, where $\Delta k=k_2-2k_1$, $k_j=n_j\omega_j/c$ are the wavenumbers and $n_j$ the linear refractive indices. Thus, $\Delta k> 0$ gives a negative (self-defocusing) cascaded nonlinearity. However, cubic nonlinearities in the material, stemming from the electronic Kerr effect, are usually positive ($n_{\rm Kerr,el}^I>0$, i.e. self-focusing), and will therefore compete with the induced self-defocusing cascading nonlinearity: to obtain a negative effective nonlinear refractive index $n_{\rm eff}^{I}=n_{\rm casc}^I+n_{\rm Kerr,el}^I$ we must have $|n_{\rm casc}^I|>n_{\rm Kerr,el}^I$. To date, \textit{critical} cascaded SHG was used for energetic pulse compression experiments \cite{liu:1999,ashihara:2002,moses:2006,ashihara:2004,*Zeng:2008,*moses:2007}. Despite not using the largest quadratic tensor components, in critical SHG the FW and SH have orthogonal polarizations and $n_{\rm eff}^{I}<0$ can be achieved by angle tuning the crystal close to the phase-matching point \cite{moses:2007a}. Consider now the normalized frequency response of the cascaded nonlinearity $R_{\rm casc}(\Omega)\equiv \Delta k/ (\tfrac{1}{2}\kpp_2\Omega^2-d_{12}\Omega+\Delta k)$ \cite{bache:2007a,*bache:2008}, where $d_{12}$ is the GVM and $\kpp_2=\tfrac{d^2k_2}{d\omega^2}$ is the SH group-velocity dispersion (GVD). Here for simplicity only up to second-order dispersion is included, making it possible to derive a threshold where $R_{\rm casc}$ becomes resonant \cite{bache:2007a,*bache:2008}
\begin{equation}\label{eq:sr}
\Delta k_{\rm r}=d_{12}^2/2\kpp_2
\end{equation}
For $\kpp_2>0$, the nonresonant regime is $\Delta k>\Delta k_{\rm r}$. 
When $\Delta k$ is significantly above 
this threshold the bandwidth $\Delta_{\rm casc}=|8(\Delta k-\Delta k_{\rm r})/\kpp_2|^{\frac{1}{2}}$ can span over an octave. The equivalent temporal response $t_{\rm casc}=1/\Delta_{\rm casc}$ is subcycle making the cascading practically instantaneous. However, if $\Delta k$ is reduced to increase the cascaded nonlinear strength, the cascaded nonlinearity narrows spectrally ($\Delta k\simeq \Delta k_{\rm r}$) and then eventually becomes resonant ($\Delta k<\Delta k_{\rm r}$), making only weak pulse compression possible before pulse distortion sets in \cite{bache:2007a,*bache:2008,bache:2010}. Note that the tunability of the cascading allows for generating a self-focusing nonlinearity as well, provided that $\Delta k<0$. In this case the requirements for an octave-spanning bandwidth are anomalous dispersion of the harmonic (here $k_2^{(2)}<0$) and $\Delta k<\Delta k_{\rm r}$.

\begin{figure}[t]
\begin{center}
\includegraphics[height=3.cm]{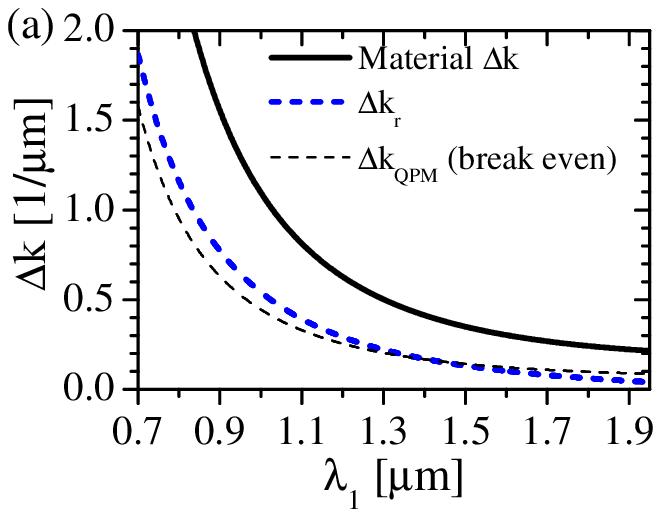}
\includegraphics[height=3.cm]{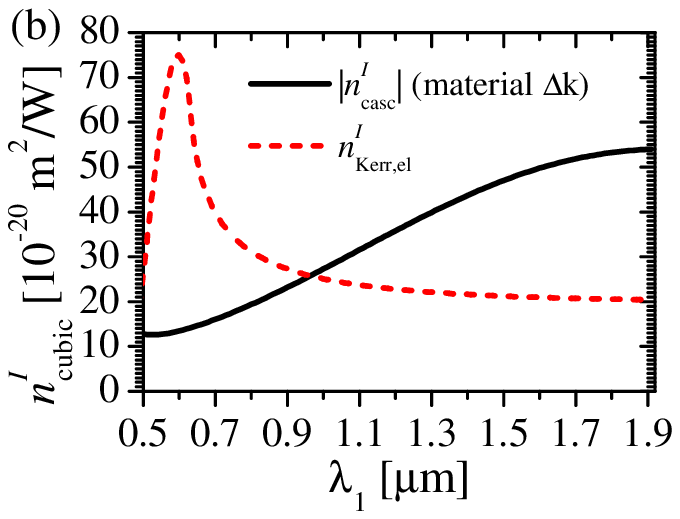}
\includegraphics[height=2.5cm]{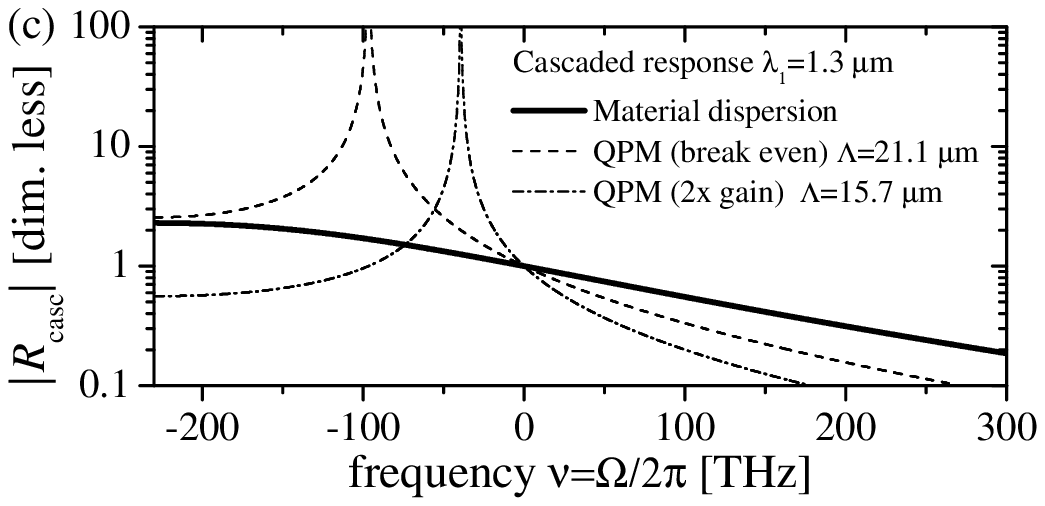}
  \end{center}
  \vspace{-6mm}
  \caption{(Color) Performance of noncritical ($\theta=\pi/2$) cascaded SHG in LN. (a) Wavelength dependence of $\Delta k$ from material-dispersion, the resonant threshold $\Delta k_{\rm r}$ from Eq. (\ref{eq:sr}), and the breakeven QPM value $\Delta k_{\rm QPM}({\rm BE})$. The Sellmeier equation from Ref. \cite{gayer:2008} was used at $T=293$ K. (b) Predicted $|n_{\rm casc}^I|$ using $d_{\rm eff}=d_{33}=-25.0$ pm/V at $\lambda=1.064\mic$ \cite{shoji:1997}, and using Miller's scaling to calculate the quadratic nonlinearity at other wavelengths, and the predicted $n_{\rm Kerr,el}^I$ \cite{Sheik-Bahae:1991} using $K=3100~\rm eV^{3/2}cm/GW$ \cite{desalvo:1996}. (c) $R_{\rm casc}$ at $\lambda=1.3\mic$ when using material-dispersion, and when using QPM to get breakeven as well as twice as  large $|n_{\rm casc}^I|$. Note: $R_{\rm casc}(\Omega)=
  \Delta k/[
  k_2(\Omega+\omega_2)-k_1^{(1)}\Omega-3k_1]$
  includes full SH dispersion.
\label{fig:n2-dk}
}
\end{figure}

Interaction through the largest quadratic tensor component promises an increased cascaded nonlinearity because $n_{\rm casc}^I\propto-d_{\rm eff}^2/\Delta k$. This makes \textit{noncritical} cascaded SHG, in which the FW and SH are polarized along the crystal axes, very attractive. Unfortunately the cascading strength is at the same time reduced due to a complete lack of phase matching. 
QPM can increase $n_{\rm casc}^I$ \cite{sundheimer:1993,*ashihara:2003a} by reducing the residual phase mismatch $\Delta k_{\rm QPM}=\Delta k-2\pi/\Lambda$, where $\Lambda$ is the QPM domain length, but since QPM also reduces $d_{\rm eff}$ (with at least $2/\pi$) the breakeven value for achieving the same $n_{\rm casc}^I$ as without QPM is $\Delta k_{\rm QPM}({\rm BE})=4\Delta k/\pi^2$. Let us investigate the consequences for lithium niobate (LiNbO$_3$, LN): Fig. \ref{fig:n2-dk}(a) shows the material $\Delta k$ vs. pump wavelength. When using QPM to reduce its large value, it always comes with the price of having $\Delta k_{\rm QPM}({\rm BE})\lesssim \Delta k_{\rm r}$. This is a consequence of a high $\Delta k_{\rm r}$ threshold caused by a large GVM that is typical of noncritical SHG. In Fig. \ref{fig:n2-dk}(c) the cascaded frequency response $R_{\rm casc}$ is plotted for $\lambda_1=1.3\mic$: when using QPM the breakeven case has a resonant cascading response as $\Delta k_{\rm QPM}({\rm BE})<\Delta k_{\rm r}$, implying that the cascading can no longer support ultrafast interaction. These properties deteriorate when QPM is forced to achieve a stronger $n_{\rm casc}^I$ than the material-dispersion case. This might have implications for ultrafast cascading using QPM, such as in Refs. \cite{Langrock:2006,*Langrock:2007}.

Since QPM seems generally unsuitable for increasing the nonlinearity, we have to use the large $\Delta k$ provided by the material. Contrary to conventional wisdom we now show that in this case the cascading properties are favorable and even advantageous. Firstly, as Fig. \ref{fig:n2-dk}(c) shows the material-dispersion case has a cascading frequency response that is nonresonant and extremely broadband (bandwidth $\Delta _{\rm casc}$ spans 1.5 octaves). This is because $\Delta k>\Delta k_{\rm r}$, and therefore the large GVM associated with noncritical interaction does not cause problems because $\Delta k$ is so big. This holds in the entire regime supporting self-defocusing solitons (up to the zero-dispersion wavelength $1.92\mic$), see Fig. \ref{fig:n2-dk}(a). Consequently, the noncritical cascaded nonlinearity is \textit{always} nonresonant and extremely broadband: we checked that its bandwidth spans over an octave in the entire plotted range and consequently the cascading is practically instantaneous (cascaded response time $t_{\rm casc}<1$ fs). Similar very favorable properties were found in all other nonlinear crystals we checked as well. The question is now whether $n_{\rm casc}^I$ is strong enough when the large material $\Delta k$ is used: Fig. \ref{fig:n2-dk}(b) shows $|n_{\rm casc}^I|$ 
whose strength increases with wavelength due to a reduced $\Delta k$. For $\lambda_1>0.96\mic$ it is larger than
$n_{\rm Kerr,el}^I$ as calculated with the two-band model \cite{Sheik-Bahae:1991}. Thus, despite the known limitations of the latter model, in this range an effective self-defocusing nonlinearity $n_{\rm eff}^I<0$ can be expected. 

\begin{figure}[t]
\begin{center}
\includegraphics[width=7cm]{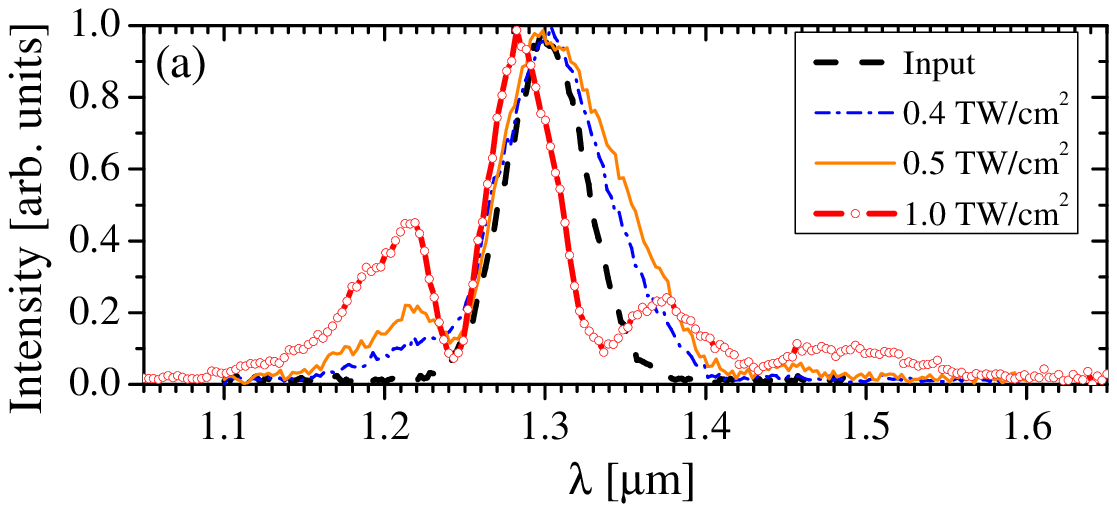}
\includegraphics[width=7cm]{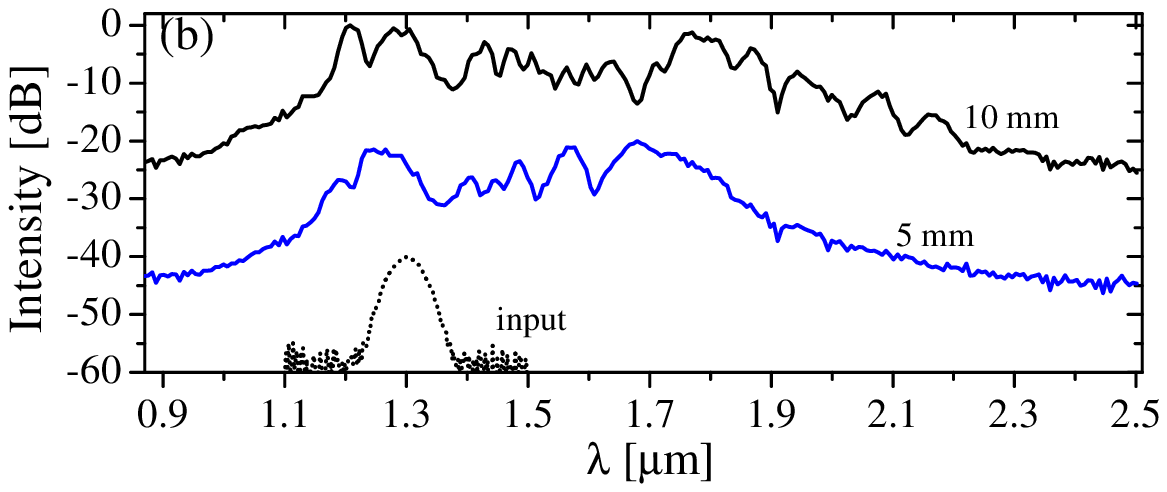}
  \end{center}
  \vspace{-6mm}
  \caption{(Color) (a) Spectra after a 1 mm LN crystal for various input intensities. (b) Spectra after 5 and 10 mm propagation using $\Iin=1.0\tw$ (averaged over 1,100 shots). The spectra in (b) are shifted 20 dB with respect to each other for clarity.
\label{fig:exp-osa}
}
\end{figure}

We conducted an experiment to illustrate the ultrafast and broadband nature of the cascading by using self-defocusing solitons to compress multicycle near-IR pulses towards few-cycle duration. A commercial Ti:Sapphire regenerative amplifier and an optical parametric amplifier generated Gaussian shaped pulses at $\lambda=1.300\mic$ with $\Delta T_{\rm FWHM}=47$ fs, 59 nm FWHM bandwidth and $\simeq 200~\mu$J pulse energy. Using a multiple-shot SHG frequency-resolved optical gating (FROG) device with a $30\mic$ thick BBO crystal the measured group-delay dispersion was $-300~\rm fs^2$, corresponding to a 42 fs FWHM transform-limited pulse. 
We found that good compression relied on ensuring that the input pulses were not positively chirped (this trend was also observed numerically). 
A broad beam spot (0.46 mm FWHM) ensured negligible diffraction, and it was controlled by a reflective silver-mirror telescope system, and the peak intensity was adjusted with a neutral density filter. The crystal was a 1-mm-long, 5\% MgO-doped $X$-cut congruent LN ($\theta=\pi/2$, $|\phi|=\pi/2$, FW and SH polarized along the vertical optical $Z$-axis, 
$10\times10~{\rm mm^2}$ aperture). The output pulses were monitored by either (a) multiple-shot intensity autocorrelator (AC) with a $100\mic$ BBO crystal, (b) FROG or (c) optical spectrum analyzer (OSA).

\begin{figure}[t]
\begin{center}
\includegraphics[width=3.cm]{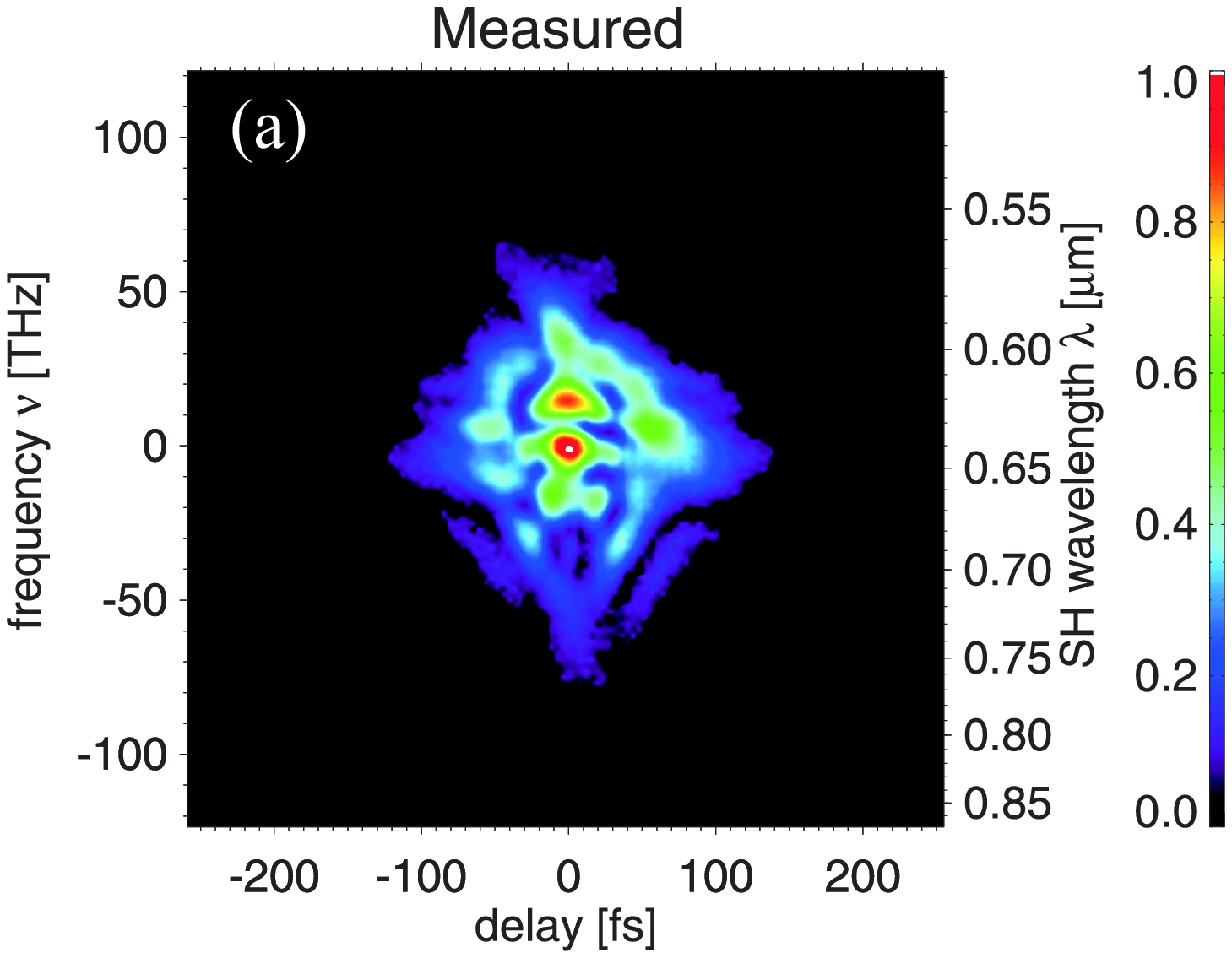}
\includegraphics[width=3.cm]{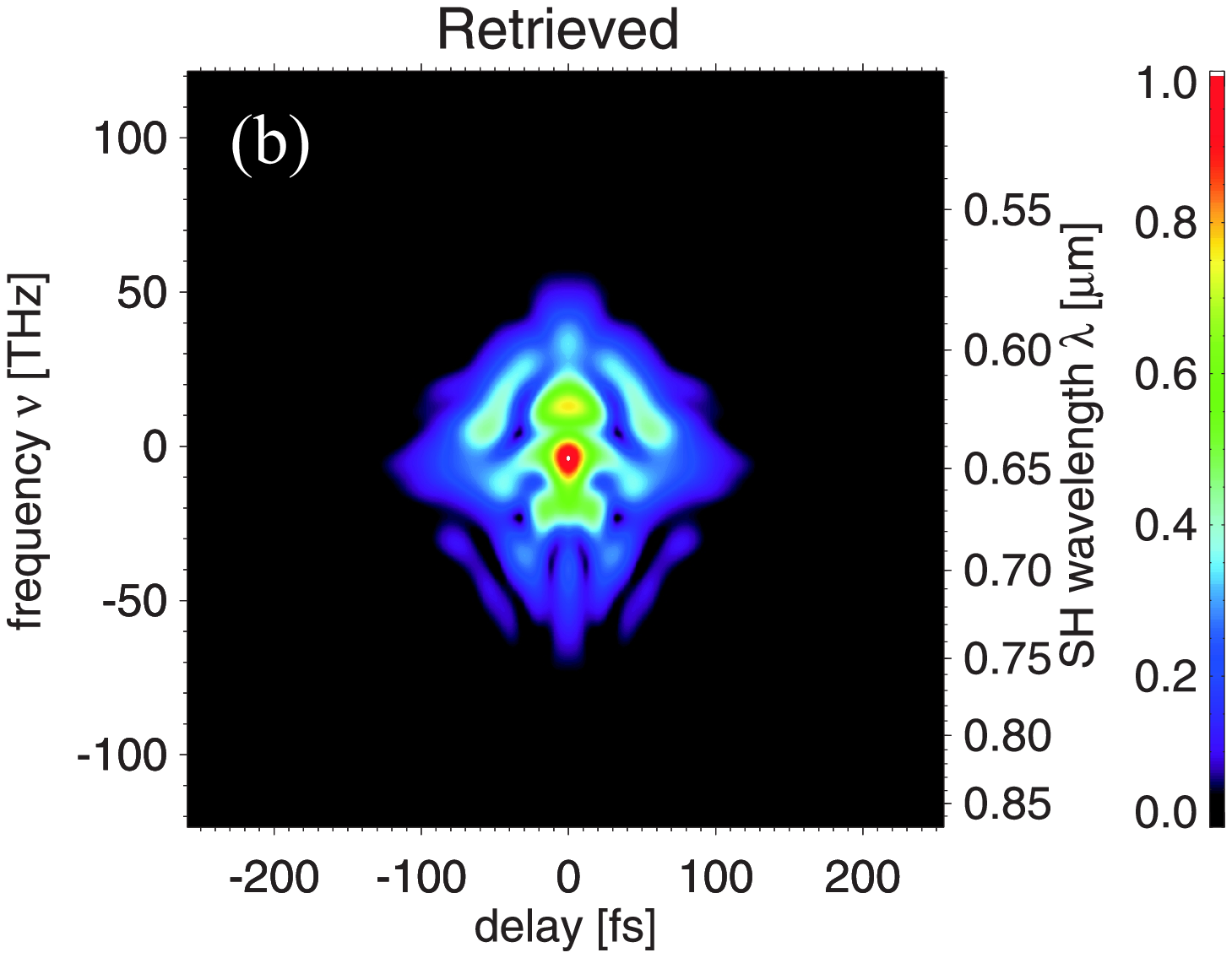}
\includegraphics[height=2.8cm]{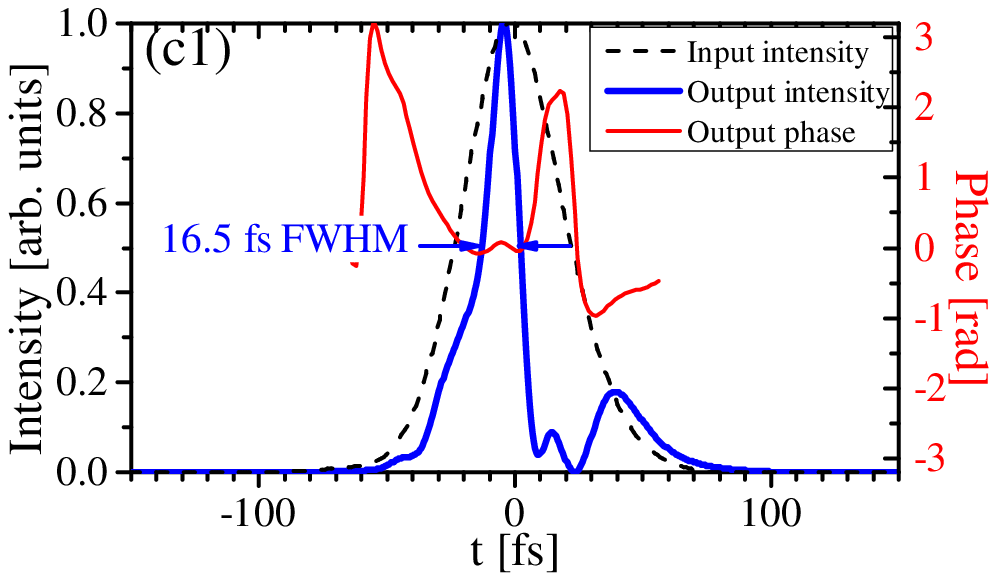}
\includegraphics[height=2.8cm]{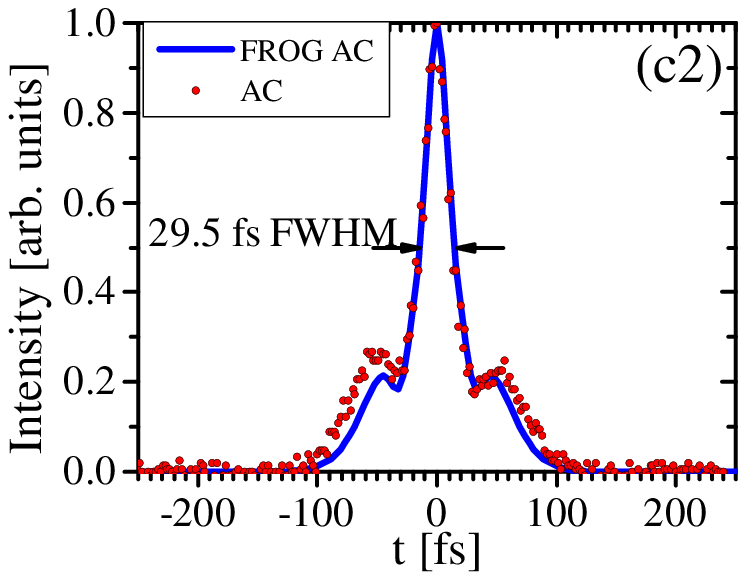}
\includegraphics[width=6.cm]{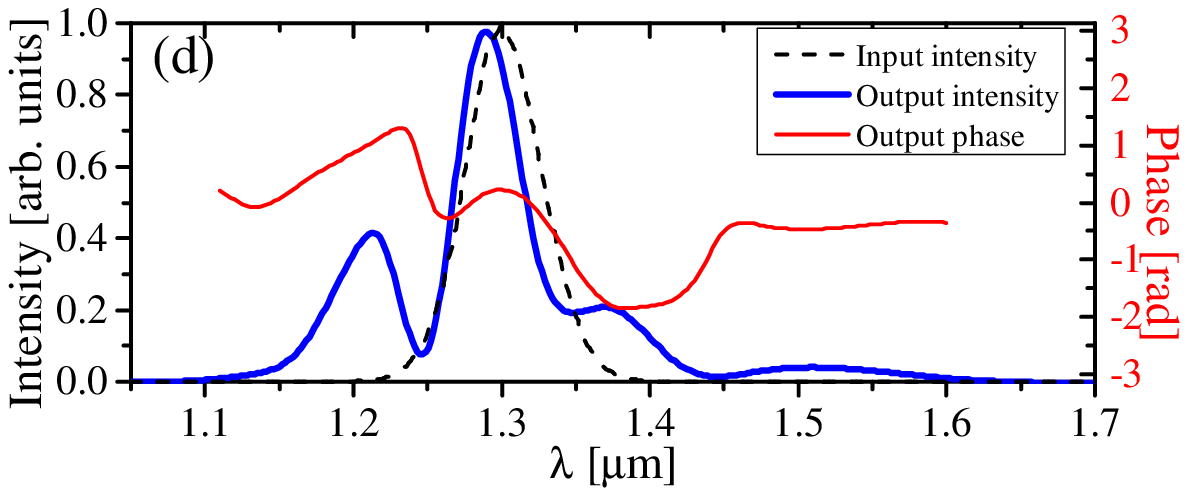}
  \end{center}
  \vspace{-6mm}
  \caption{(Color) FROG measurements at $\Iin=1.0~\rm TW/cm^2$: (a,b) FROG traces (amplitude) of the output pulse. (c1) The temporal and (d) spectral amplitudes and phases retrieved by the FROG algorithm. 
  (c2) The calculated FROG intensity AC and the experimentally measured AC.
\label{fig:exp-frog}
}
\end{figure}

Figure~\ref{fig:exp-osa}(a) shows representative spectra measured by the OSA. For lower intensities, some spectral broadening occurs. When the intensity reaches $\Iin=0.5\tw$, a weak shoulder appears on the blue side of the central lobe. At this point the AC traces evidenced pulse compression, which became strongest at $\Iin=1.0\tw$ (0.13 mJ pulse energy). 
The FROG results in Fig. \ref{fig:exp-frog} reveal more details at this intensity. The retrieved temporal intensity shows clear pulse compression, and a smaller delayed pulse. A Gaussian fit to the central spike gave 16.5 fs FWHM (below 4 optical cycles), and by comparing the fitted-curve area with the total area, we estimate an efficiency of 80 \%. The flat phase in the time trace indicates that we are close to the soliton formation point. The spectral phase is also quite flat and indeed by calculating the transform limited pulse supported by the spectrum gave a 14 fs FWHM pulse, close to the observed value. The retrieved FROG spectral intensity in Fig. \ref{fig:exp-frog}(d) is in excellent agreement with the OSA spectrum in  Fig.~\ref{fig:exp-osa}, and similarly the FROG AC trace agrees well with the independently measured AC trace, see Fig. \ref{fig:exp-frog}(c). 

Below $1.92\mic$ LN is normally dispersive (GVD coefficient $\kpp_1=+162~{\rm fs^2/mm}$ at $1.3\mic$), so the combination of spectral broadening and temporal compression well below the transform limit of the input pulse proves that we have excited a self-defocusing soliton. 
In fact, if the nonlinearity were self-focusing, no pulse compression could be achieved as such a nonlinearity would require anomalous GVD compensation after the crystal. 
We checked with an IR camera that the FW showed no diffraction, which can be attributed to the short crystal and broad spot size. Soliton formation in such a short crystal is a consequence of the huge FW GVD associated with using the largest nonlinear tensor components, which implies that only a short propagation distance is needed to compensate for the nonlinear SPM-induced chirp.


Tuning the intensity around the value for optimal compression, $\Iin=1.0\tw$, the experiment showed similar pulse compression and spectral traces in the range $\Iin=0.8-1.3\tw$. Beyond $1.3\tw$ substantially more complex autocorrelation traces were observed. For smaller intensities the temporal compression gradually disappeared. This is typical of soliton compression and is consistent with numerical simulations. 

With further propagation, the spectrum continues to broaden rapidly; the physics behind this is similar to standard supercontinuum generation \cite{dudley:2006}, namely soliton fission after the initial compression point followed by various nonlinear interaction steps that broaden the spectrum. The generation of an octave-spanning supercontinuum spectrum [Fig. \ref{fig:exp-osa}(b)] in longer crystals is direct evidence of the ultrabroad bandwidth of the cascading process we investigate here.

\begin{figure}[t]
\begin{center}
\includegraphics[width=8.5cm]{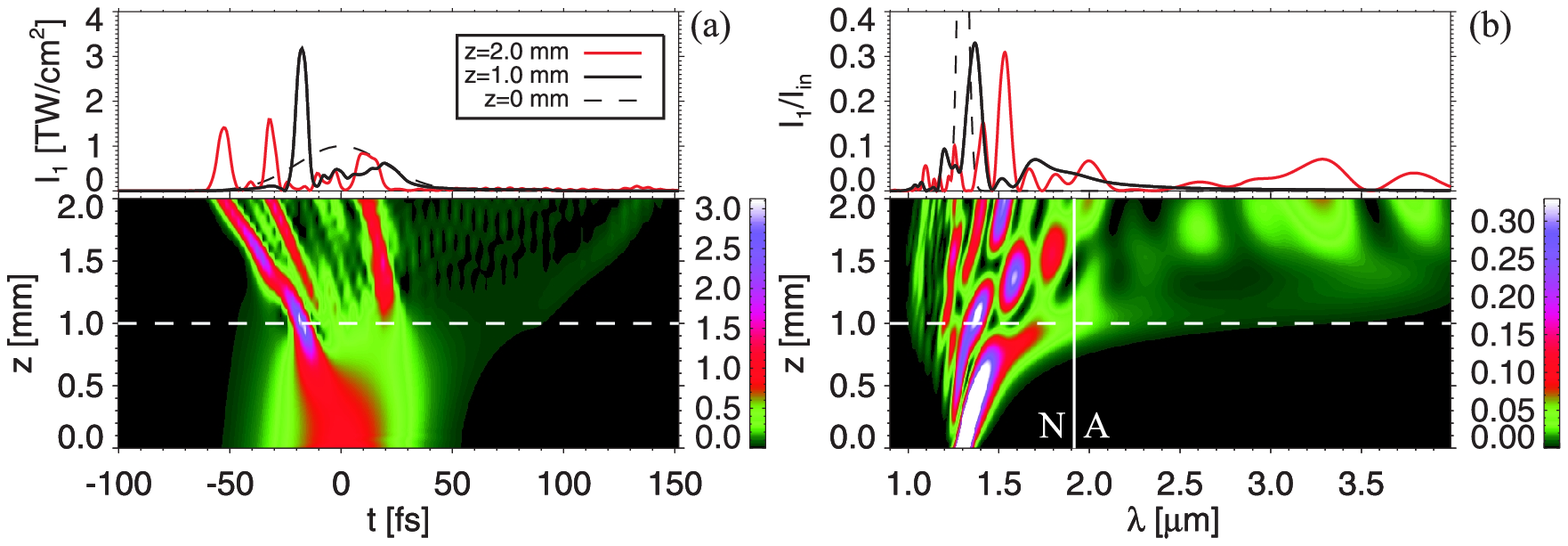}
\includegraphics[width=8.5cm]{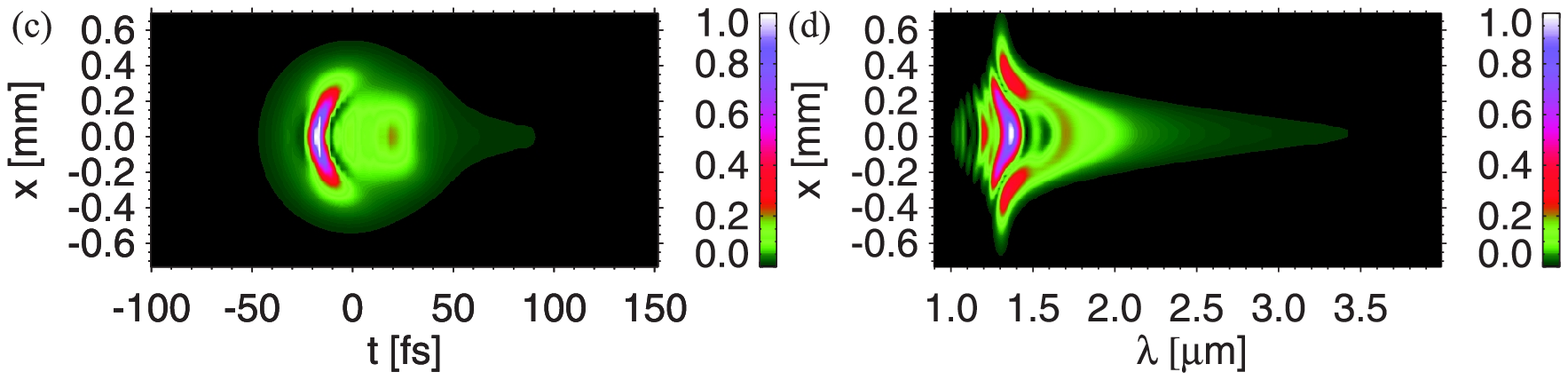}
  \end{center}
  \vspace{-6mm}
  \caption{(Color) Numerical simulation of 2 mm propagation in LN with $\Iin=1.0\tw$. The FW on-axis evolution along $z$ is shown as (a) time intensity and (b) spectral intensity. 'N' and 'A' indicate the normal and anomalous dispersion regimes, and the color scale is saturated for clarity. The top cuts show the intensities at input, $z=1$ mm (the position indicated with a dashed white line) and $z=2$ mm. (c) and (d) show the transverse variation along the $x$ direction of the time and spectral contents at $z=1$ mm and fixing $y=0$. 
\label{fig:sim}
}
\end{figure}

We performed numerical simulations using the 3+1D slowly evolving wave approximation \cite{moses:2006b,*bache:2007,*Bache:2007-erratum}, including the Raman effect modeled as a 4-mode complex Lorentzian in frequency domain (for Raman spectrum parameters, see \cite{Kaminow:1967,*Kaminow:1969}). Based on Raman peak-gain \cite{Kaminow:1967,*Kaminow:1969} and $Z$-scan measurements \cite{desalvo:1996}, we made a detailed analysis (which will be published elsewhere) and found that at $\lambda=1.3\mic$ the experimental results \cite{desalvo:1996,Kaminow:1967,*Kaminow:1969} support the values $n_{\rm Kerr}^I=45\times 10^{-20}~\rm m^2/W$ and $f_R=0.50$ (both $\pm10\%$), giving $n_{\rm Kerr,el}^I=(1-f_R)n_{\rm Kerr}^I=22.5\times 10^{-20}~\rm m^2/W$. 
In Fig. \ref{fig:sim} we used the same input pulse parameters as the $\Iin=1.0\tw$ experiment. The on-axis spectrum (b) shows a strong central lobe, a dominant blue shoulder and a weaker red shoulder followed by a broad plateau. The on-axis time trace (a) also shows a strong spike and a weaker trailing pulse. The main soliton is shorter than what we observed experimentally, but the simulation indicates that the trailing pulse we saw is a weaker soliton formed by soliton fission (the intensity corresponds to an effective soliton order of around 6 for this simulation). We found that this kind of early-stage soliton fission was mainly caused by the rather large soliton order combined with strong Raman effects. This fission process continues so after 2 mm three solitons exist, and upon further propagation an extremely wide supercontinuum is formed. 
These simulations also confirmed that no diffraction occurs, as was observed in the experiment. Moreover, concerning the spatio-temporal distribution of the compressed soliton, the space-time cross-section in Fig. \ref{fig:sim}(c) shows that it is quite homogeneous (similar to the Gaussian case of Ref. \cite{moses:2007}). 
Finally, note the optical Cherenkov waves that emerge in Fig. \ref{fig:sim} as broadband waves centered around $\lambda=3.0\mic$; such waves could be an efficient source of mid-IR few-cycle radiation  \cite{bache:2010e,*bache:2011a}.
The simulations indicate the possibility of exciting self-defocusing solitons at shorter wavelengths as well. At some point, however, the effective nonlinearity will become focusing, see Fig. \ref{fig:n2-dk}(b). At longer wavelengths the Kerr, and in particular the Raman, nonlinearities will be less dominating, making cascading more favorable. 

Concluding, strongly phase-mismatched cascading exploiting the largest quadratic nonlinear tensor element can create ultrafast and octave-spanning self-defocusing cascaded nonlinearities. We confirmed this by the first experimental observation of few-cycle self-defocusing soliton pulse compression in noncritical cascaded SHG, which occurred in a short (1 mm) bulk lithium niobate crystal. Upon further propagation an octave-spanning supercontinuum was generated. The noncritical interaction gives short interaction lengths and zero spatial walk-off, which represent significant improvements compared to previous experiments 
\cite{liu:1999,ashihara:2002,moses:2006,ashihara:2004,*Zeng:2008,*moses:2007}. Because the nonlinearity is self-defocusing the pulse energy is readily scalable in large-aperture crystals to multi-millijoules. As it is also compact, cheap and works already for few-microjoule pulses, this compressor may replace or fill out gaps of standard hollow-fiber compressors \cite{DeSilvestri:2004}.


Paradoxically, \textit{despite} the strong phase mismatch a strong self-defocusing cascaded nonlinearity is achieved even without QPM owing to a huge quadratic nonlinearity. However, it is \textit{exactly} the strong phase mismatch that ensures a nonresonant octave-spanning cascaded nonlinearity, which is practically instantaneous, so even with a large group-velocity mismatch ultrafast femtosecond interaction is possible.
This discovery may open a new era of ultrafast cascading based on strongly phase-mismatched interaction exploiting the largest quadratic nonlinearities, in particular with semiconductor materials that are known for their huge quadratic nonlinearities that cannot be phase matched. It also implies that QPM is not needed to increase the cascaded nonlinearity. Thus, some of the technological limitations of QPM are removed, like complexity in production, poling quality, limited beam aperture etc. Not only that, but avoiding QPM is often crucial for ultrafast purposes as the cascaded nonlinearity can become resonant with QPM.


Acknowledgments: Jeffrey Moses, 
Xianglong Zeng for comments and discussions; the Danish Council for Independent Research (\htmladdnormallink{\textit{Femto-VINIR}}{http://www.femto-vinir.fotonik.dtu.dk} project, 274-08-0479); the US National Science Foundation (PHY-0653482).



%

\end{document}